\documentclass[twocolumn]{jpsj2}

\def\runauthor{Online-Journal Subcommittee}

\usepackage{graphicx}
\usepackage{dcolumn}
\usepackage{bm}

\title{Phase diagram of orbital-selective Mott transitions
at finite temperatures}

\author{Kensuke {\sc Inaba},
Akihisa {\sc Koga},
Sei-ichiro {\sc Suga} and
Norio {\sc Kawakami}}

\inst{%
Department of Applied Physics, Osaka University, Suita, Osaka 565-0871 
}%

\recdate{\today}

\abst{
 Mott transitions in the two-orbital Hubbard model with different 
bandwidths are investigated at finite temperatures.
By means of the self-energy functional approach, we discuss the 
stability of the intermediate phase with one orbital localized 
and the other itinerant, 
which is caused by the orbital-selective Mott transition (OSMT).
It is shown that the OSMT realizes two different coexistence regions
at finite temperatures in accordance with the recent results of
Liebsch.
We further find that the particularly interesting behavior
emerges around the special condition $U=U'$ and $J=0$,
which includes a new type of the coexistence region with three
distinct states.  By systematically changing the 
Hund coupling, we establish the global phase diagram to
elucidate the key role played by the Hund coupling on the 
 Mott transitions.
}

\kword{Orbital-selective Mott transition, Self-energy functional approach}

\begin{document}
\maketitle


Strongly correlated electron systems with some orbitals have
 been investigated  extensively.\cite{ImadaRev,TokuraScience}
In particular, substantial progress in theoretical 
understanding of Mott transitions of multi-orbital systems 
has been made by dynamical mean-field theory (DMFT) \cite{GeorgesRev,KotliarPT} calculations. \cite{Kim90,Kotliar96,Rozenberg97,Bunemann:Gutzwiller,Hasegawa98,Held98,Han98,Momoi98,Klejnberg98,Imai01,Koga:ED_DMFT,Oudovenko02,Ono03,Tomio04,Pruschke04,Sakai04,Liebsch,Koga04,Koga05,Ferrero05,Medici05,Arita05,Knecht05,note,Biermann05}
Among them, the orbital-selective Mott transition (OSMT)\cite{Anisimov02}
in the multi-orbital system with different bandwidths has been
one of the most active topics in this context. 
Typical materials are 
$\rm Ca_{2-x}Sr_xRuO_4$\cite{Nakatsuji} and 
$\rm La_{n+1}Ni_nO_{3n+1}$\cite{LaNiO,Kobayashi96}, where
the OSMT is suggested to be realized by the chemical substitution 
and the change in the temperature.
These experimental findings have stimulated theoretical 
investigations on this issue.
\cite{Liebsch,Sigrist04,Fang04,Anisimov02,Koga04,Koga05,Ferrero05,Medici05,Arita05,Knecht05,note,Biermann05}
 It has been clarified 
by extensive works that in the two-orbital system with different
bandwidths at zero temperature, the OSMT occurs in general and merges to
a single Mott transition in a certain restricted parameter region.
\cite{Koga04,Ferrero05,Medici05}
 Also,
it has been realized that the Hund coupling is an important key parameter
 to understand the correct nature of the OSMT.\cite{KogaPr,note}

In contrast to the detailed analysis at zero temperature, 
systematic discussions at finite temperatures
\cite{Koga05,Liebsch,note,Knecht05} 
are lacking. 
In particular, the stability of the intermediate metallic phase 
may be important to discuss the possibility of the OSMT in real
 materials.  It is thus desirable to investigate the OSMT
in the multi-orbital model at finite temperatures.
Quite recently, Liebsch have studied  finite-temperature properties 
of the OSMT by DMFT combined with 
the exact diagonalization method,\cite{note} where the OSMT
occurs even at finite temperatures.

In this paper, we give the detailed investigation of 
the OSMT in the two-orbital Hubbard model at zero and finite temperatures.
We make use of the self-energy functional approach (SFA) proposed by Potthoff,\cite{SFA} 
which is efficient to study finite-temperature properties, 
to determine the detailed phase diagram.
The present study provides the results complementary to
Liebsch's in a certain parameter regime,\cite{note} and moreover establishes
rich phase diagrams, which include a new type of the 
coexistence region with three distinct states.
We further clarify how the magnitude of the Hund coupling controls
 the nature of phase diagrams at finite temperatures.



We consider the two-orbital Hubbard  model with different bandwidths,
which is given by the 
Hamiltonian, ${\cal H}={\cal H}_0+\sum_i {\cal H}_i^\prime$ with
\begin{eqnarray}
{\cal H}_0&=&\sum_{<i,j>, \alpha, \sigma}
\left( t_\alpha -\mu\delta_{ij}\right)
          c^\dag_{i\alpha\sigma} c_{j\alpha\sigma},\\
{\cal H}_i^\prime&=& \!\!\!\!
                  U \sum_{\alpha}
                n_{i \alpha \uparrow} n_{i \alpha \downarrow}
                  +\sum_{\sigma \sigma^\prime}
                 (U^\prime - \delta_{\sigma\sigma^\prime}J)
                n_{i 1 \sigma} n_{i 2 \sigma^\prime}
                \nonumber\\
          &-&  \!\!\!\!\! J  (c^\dag_{i 1 \uparrow}c_{i 1 \downarrow}
                c^\dag_{i 2 \downarrow}c_{i 2 \uparrow}
                +c^\dag_{i 1 \uparrow}c^\dag_{i 1 \downarrow}
                c_{i 2 \uparrow}c_{i 2 \downarrow}+H.c.)
          \label{eq:model},
\end{eqnarray}
where $c^\dag_{i\alpha\sigma}(c_{i\alpha\sigma})$ is 
the creation (annihilation) operator of an electron at the $i$th site
 with spin $\sigma(=\uparrow,\downarrow)$ and orbital $\alpha(=1,2)$, 
and $n_{i\alpha\sigma}=c^\dag_{i\alpha\sigma}c_{i\alpha\sigma}$.
Here, $t_\alpha$ denotes the hopping integral for orbital $\alpha$, 
$\mu$ the chemical potential, 
$U (U')$ the intra-orbital (inter-orbital) Coulomb interaction, 
and $J$ the Hund coupling including the spin-flip and pair-hopping terms. 
In the following, we impose the condition $U=U'+2J$,
which results from rotational symmetry of degenerate orbitals.


In order to discuss the Mott transitions at zero and finite temperatures,
we use here the self-energy functional approach SFA,\cite{SFA}
which is based on the Luttinger-Ward variational method.\cite{Luttinger}
This approach allows us to deal with 
finite-temperature properties of the multi-orbital system efficiently,
\cite{Inaba05} 
where standard DMFT with numerical techniques
may encounter some difficulties in a practical computation when 
the number of orbitals increases. 

In SFA, one makes use of the fact that the Luttinger-Ward functional does not depend on 
the detail of the Hamiltonian ${\cal H}_0$ as far as the interaction term
${\cal H}^\prime$ is unchanged.\cite{SFA}
This enables us to introduce a proper reference system with
the same interaction term.
One of the simplest models for the reference system is explicitly given 
by the following Hamiltonian,
${\cal H}_{\rm ref}=\sum_i{\cal H}_{\rm ref}^{(i)}$, 
\begin{eqnarray}
  {\cal H}_{\rm ref}^{(i)}&=&\sum_{\alpha \sigma } \left[ 
\epsilon^{(i)}_{0\alpha}
          c^\dag_{i\alpha\sigma} c_{i\alpha\sigma}+          
\epsilon^{(i)}_{\alpha}
          a^{(i)\dag}_{\alpha\sigma}a^{(i)}_{\alpha\sigma}\right]\nonumber\\
          &+&\sum_{\alpha \sigma}V^{(i)}_{\alpha}
          (c^\dag_{i\alpha\sigma}a^{(i)}_{\alpha\sigma}+H.c.)
+H_i^\prime,\label{eq:ref_model}
\end{eqnarray}
where $a^{(i)\dag}_{\alpha\sigma}(a^{(i)}_{\alpha\sigma})$ 
creates (annihilates) an electron with $\sigma$ spin and $\alpha$ orbital, 
which is connected to the $i$th site in the original lattice. 
This approximation is regarded as a finite-temperature extension
of the two-site DMFT\cite{Potthoff01} in some respects.
In this paper, we fix the parameters 
$\epsilon_{0\alpha}=0, \epsilon_{\alpha}=\mu$ and $\mu=U/2+U'-J/2$ 
to discuss the zero and finite temperature properties at half filling.
By choosing the parameters $V_{\alpha}$ to minimize the grand potential, 
$\partial\Omega/\partial V_\alpha=0$ $(\alpha=1,2)$,
we can find a proper reference Hamiltonian,
 which approximately describes the original correlated system.

We note that the hybridization $V_\alpha$ between a given site on the
original lattice and a site in the reference system may be regarded as 
the renormalized bandwidth of the $\alpha$-orbital band.
For example, for small $V_\alpha$, heavy quasi-particles are formed 
in the $\alpha$ band, and at $V_\alpha =0$ the system 
is driven to the Mott insulating state.
By calculating hybridizations $V_1$ and $V_2$, 
we thus discuss the stability of the metallic state in 
our two-orbital Hubbard model.
In the following, we focus on the system with $W_1=2.0$ and $W_2=4.0$ 
to discuss Mott transitions, 
where $W_\alpha$ is the bandwidth of $\alpha$-orbital 
for the semi-circular density of states (DOS),
$\rho_\alpha(x)=4/\pi W_\alpha \sqrt{1-(2x/W_\alpha)^2}$.

\begin{figure}[htb]
\includegraphics[width=\linewidth]{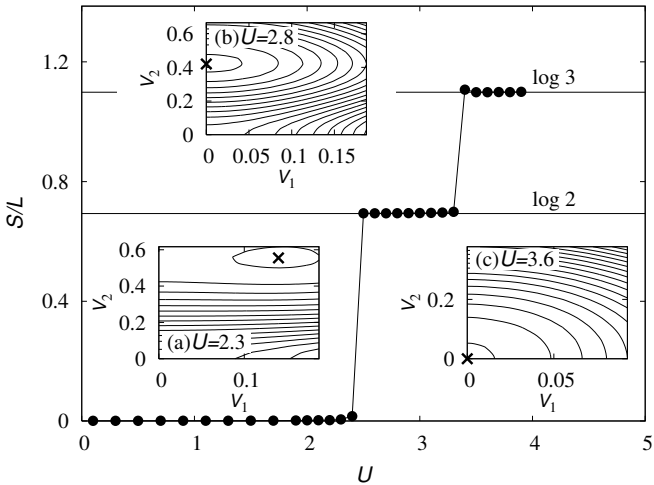}
\caption{
The entropy per site $S/L$ as a function of $U$ with fixed ratios 
$U'=0.5U$ and $J=0.25U$ at $T=0$.
Inset shows the contour plot of the grand potential 
for (a)$U=2.3$, (b) 2.8 and (c) 3.6, where crosses denote the minima 
of the grand potential.}
\label{fig:zero:ent}
\end{figure}

We begin our discussions with the Mott transitions 
at zero temperature for a typical choice of
the ratio of the parameters, $U'=0.5U$ and $J=0.25U$.
 Figure \ref{fig:zero:ent} shows the contour plots of 
the grand potential  and the corresponding
entropy per site $S/L$ at the stationary point, where
$L$ is the number of sites.
When the Coulomb interaction is small $(U=2.3)$, 
the stationary point, where the grand potential is minimized, 
is located around 
$(V_1, V_2)\sim (0.14, 0.55)$.
This implies that the effective bandwidth for each orbital state is 
finite,  stabilizing the metallic (M) phase. 
Increasing the interaction, the stationary point of the system 
touches the $V_2$-axis as shown in the inset (b):
the effective bandwidth for the narrower band $V_1$ vanishes while
$V_2$ still remains finite.
Therefore, the OSMT occurs, where
one band is insulating  and the other is still metallic.
 We refer to 
this phase as the OSM phase in the 
following. It is seen that the residual entropy 
in the OSM phase is $S/L = \log 2$,  implying the formation of
free localized  $s=1/2$ spins.
Further increase of the interaction induces the Mott transition 
in the other band to the Mott insulating (MI) phase,
where $V_1=V_2=0$ as shown in the inset (c).
In this case, the localized spins for two bands form the triplet 
state due to the Hund coupling, so that the residual entropy $S/L=\log 3$ 
 in the region $U>3.4$.  Note that the residual entropy at zero temperature 
is an artifact of our simplified SFA approach. This pathological behavior can be
somehow improved, e.g. by taking into account spin fluctuations
more accurately.
At zero temperature, we can find no other local minima 
in the grand potential when the interaction is varied.
Therefore, we clearly observe double second-order Mott 
transitions in the multi-orbital system, which indeed confirms 
the claim of Koga {\it et al}.\cite{Koga04} for
the OSM transition obtained
 by means of DMFT combined with the exact diagonalization.

\begin{figure}[ht]
\includegraphics[width=.9\linewidth]{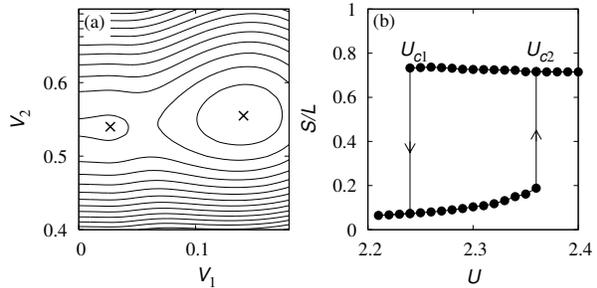}
\caption{(a) The contour plot of the grand potential $\Omega(V_1,V_2)$ 
at $T=0.002,U=2.3$ and  $J=0.25U$.  Two minima coexist, which correspond to 
 the M and OSM states. 
(b) The entropy vs the interaction $U$ at $T=0.002, J=0.25U$.
}
\label{fig:local}
\end{figure}

\begin{figure}[ht]
\includegraphics[width=\linewidth]{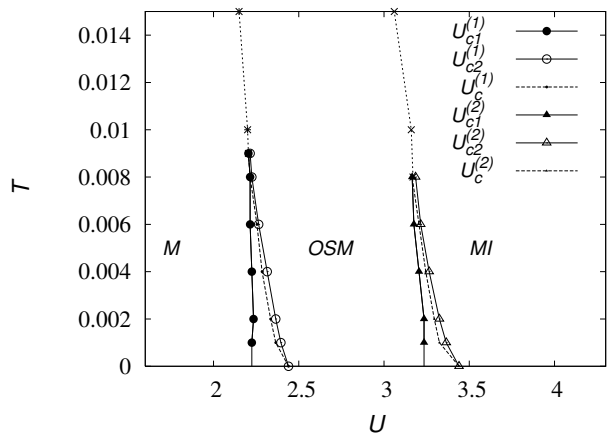}
\caption{$U-T$ phase diagram at $J=0.25U$. 
There are two coexistence phases, which are featured by the 
triangular-shaped regions: ({\it left})the metallic phase and 
orbital-selective Mott phase coexist  and ({\it right}) orbital-selective 
Mott phase and insulating phase  coexist.}
\label{fig:phase_J25}
\end{figure}

\begin{figure}[t]
\includegraphics[width=\linewidth]{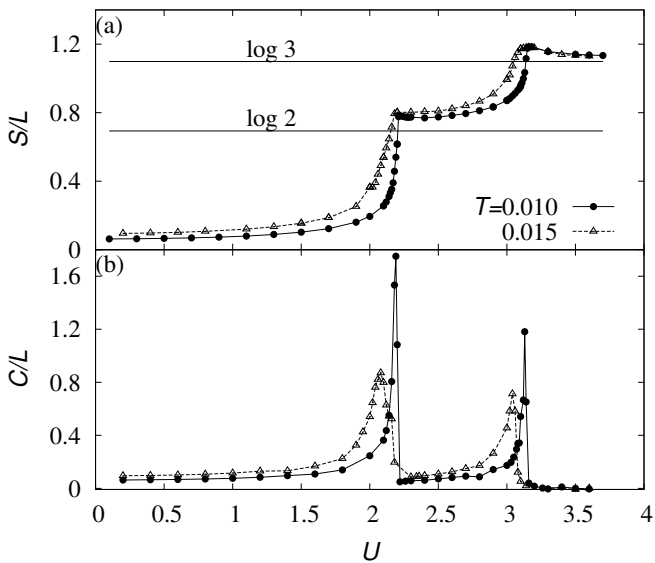}
\caption{(a) Entropy $S/L$ and (b) Specific heat $C/L$
as a function of $U$ in the crossover region, $J=0.25U$.
}\label{fig:hcap}
\end{figure}

We now move to the phase diagram at finite temperatures.
Since each transition at zero
temperature is similar to that for the single-orbital Hubbard model,
\cite{GeorgesRev}
the first-order transition may occur at finite temperatures 
around each critical point, as shown in ref. \citen{note}.
In fact, we find that when the system belongs to the metallic phase 
close to the critical point, two stationary points representing the M and 
OSM states appear simultaneously once the system is at finite temperatures
 ($T=0.002$), as shown in Fig. \ref{fig:local}.
The double-well structure causes
the first-order Mott transition with hysteresis at finite temperatures
(Fig. \ref{fig:local} (b)). At $T=0.002$, as $U$ increases, 
the stationary point for the metallic
 state disappears around $U_{c2}\sim 2.36$, 
where the Mott transition occurs to the OSM phase.
On the other hand, as $U$ decreases, the OSM phase realized is stable 
down to
$U_{c1}\sim 2.24$.
The first-order transition point $U_c\sim 2.33$ for $T=0.002$ is determined 
by the crossing point of the two minima in the grand potential.
The obtained phase diagram is shown in Fig. \ref{fig:phase_J25}.
It is found that the two distinct coexistence regions 
appear around $U\sim 2.4$ and $U\sim 3.3$.
The phase boundaries $U_{c1}$, $U_{c2}$ and $U_c$ merge 
at the critical temperature $T_c$ for each transition.
  Similar phase diagram 
was recently obtained by Liebsch by means of DMFT with 
the exact diagonalization method, who gave better estimates 
of the critical temperatures.\cite{note} Nevertheless, our SFA treatment 
elucidates further interesting 
properties such as the crossover behavior among the competing 
M, OSM and MI phases.  To make this point more explicit,
we calculate thermodynamic quantities
such as the entropy and the specific heat, which are 
displayed in Fig. \ref{fig:hcap}.
We can see a double-step structure 
similar to that found at zero temperature
in the curve of the entropy. Such anomalies 
 are more clearly seen in  the specific heat.  It is quite impressive 
that the crossover behavior among three phases are clearly seen
even at higher temperatures. We can thus say that
the OSM phase is rather well defined even at higher temperatures
 above the critical temperatures.



\begin{figure}[ht]
\includegraphics[width=\linewidth]{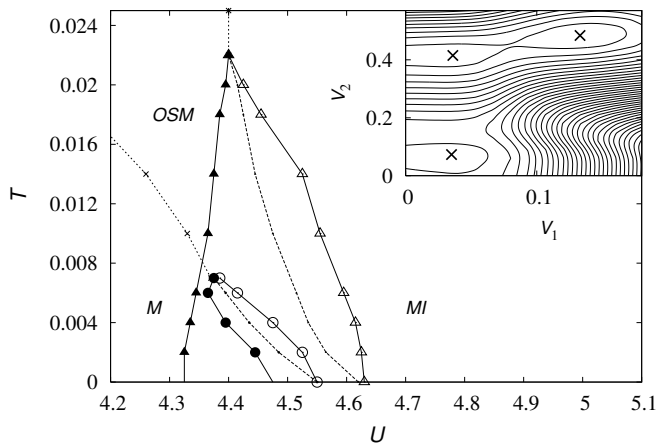}
\caption{$U-T$ phase diagram at $J=0.03U$: we have used the
same  symbols as in Fig. \ref{fig:phase_J25}.
Note that the two coexistence regions are not separated here in
contrast to Fig. \ref{fig:phase_J25}. 
Three distinct states (M,OSM,and MI states) coexist
in the region which is surrounded by filled and open circles.
Inset shows the contour plot of the ground potential in the
 coexistence region for $U=4.45,T=0.004$, where there are  three 
local minima. 
}\label{fig:phase_J3}
\end{figure}

We have so far discussed  the OSMT
for the system with rather large  Hund couplings. 
According to the zero-temperature analysis,\cite{Koga04,Ferrero05,Medici05} 
it is known that around
the special condition $U=U'$ and $J=0$ at $W_1/W_2=0.5$, 
a non-trivial single Mott transition occurs, whose origin is
attributed to enhanced orbital fluctuations.\cite{Koga05}
Therefore, it is interesting to observe what happens 
at finite temperatures,  when the system approaches the 
condition $U=U'$ and $J=0$.
By performing similar calculations, we end up with the phase diagram for 
the system with  $U'=0.94U, J=0.03U$  (Fig. \ref{fig:phase_J3}).
Since the Hund coupling is very small in this  case, 
two critical points for the OSMT
at zero temperature are close to each other.
One of the most remarkable findings here is that 
three competing states (M, OSM and MI states)
coexist in a certain region at finite temperatures, 
where three distinct local minima appear 
in the grand potential, as shown in the inset.
In these parameters,
 the coexistence regions are quite  sensitive to how large 
the Hund coupling is. 

\begin{figure}[t]
\includegraphics[width=.9\linewidth]{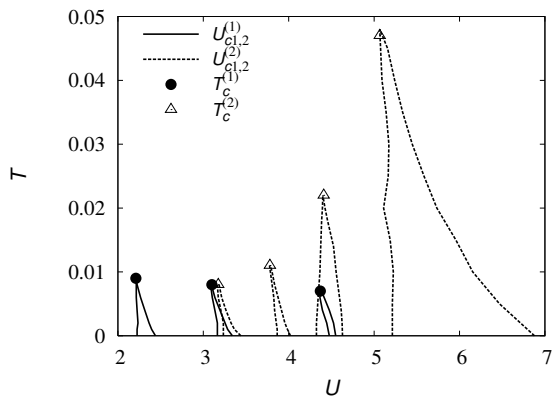}
\caption{Overall picture for the coexistence regions with 
systematic change of  $J$: two types of the regions 
 are shown for $J=0.25U, J=0.1U$, $J=0.03U$ and $J=0$ from 
left to right.
At $J=0$, the transition from the metal to OSM disappears.
}\label{fig:phase_all}
\end{figure}

To observe how the Hund coupling controls finite-temperature 
properties, we show the global phase 
diagrams in Fig. \ref{fig:phase_all} by systematically changing $J$.
As the Hund coupling  decreases, 
 both the critical points $U_{c2}^{(1)}$ and $U_{c2}^{(2)}$ 
increase monotonically, stabilizing the metallic state
 up to fairly large Coulomb interactions, as already 
mentioned for the zero-temperature case.\cite{Koga04} A remarkable feature 
 in Fig. \ref{fig:phase_all} is that the completely different 
behavior emerges in the critical temperatures $T_c^{(1)}$ and $T_c^{(2)}$.
It is seen that $T_c^{(1)}$ is almost
 unchanged, whereas $T_c^{(2)}$ strongly depends on the Hund coupling. 
This may be explained as follows. Around the first 
boundary between the M and OSM phases, the
 nature of the transition is essentially the same as 
the single-orbital case, as discussed above. Therefore, 
the Hund coupling little affects the critical temperature
$T_c^{(1)}$. On the other hand, around 
the second phase transition between the OSM and 
MI phases, orbital fluctuations should play a vital role. 
The presence of the Hund coupling somewhat suppresses 
orbital fluctuations due to the formation of the triplet state.\cite{Koga:ED_DMFT}
Therefore,  as $J$ decreases, enhanced orbital fluctuations raise 
the critical temperature $T_{c}^{(2)}$, as seen in Fig. 
\ref{fig:phase_all}.
Note that for very small Hund couplings
$(J\lesssim 0.03U)$  the OSM phase disappears,\cite{Medici05} and thus
the finite as well as zero-temperature properties 
can be approximately described by  the multi-orbital model 
$(U\neq U')$ with the same bandwidths. 
\cite{Koga:ED_DMFT,Inaba05,Ono03,Pruschke04}

In summary, we have investigated Mott transitions in the 
two-orbital Hubbard model with different bandwidths at finite 
temperatures. By means of the self-energy functional approach,
we have confirmed that two distinct coexistence regions appear 
in the phase diagram in accordance with the recent results obtained
by DMFT. A notable new finding is that the OSM phase is rather well
defined even at temperatures  higher than the two critical 
temperatures. Further remarkable features have been elucidated in the
phase diagram around special conditions with small Hund couplings
$J \simeq 0$ ($U \simeq U'$), where the coexistence region 
with three competing phases emerges. Orbital 
fluctuations are enhanced there,
and therefore the system gets very sensitive to
small perturbations, giving rise to the rich 
phase diagram at low temperatures.
We have also clarified the role of the Hund coupling 
on global features in the phase diagram: the Hund coupling has 
little (sizable) effect on the transition between the M and OSM 
 (OSM and MI) phases, which is again attributed to the role played by 
enhanced orbital fluctuations. 

In this paper we have exploited the simplified version of SFA to
elucidate fundamental properties at finite temperatures. Since 
instabilities to possible ordered phases have not been considered here, 
 SFA is to be generalized to incorporate such ordered phases, which 
should be done in the future work. 

We would like to thank M. Sigrist, T. M. Rice 
and A. Liebsch for valuable discussions.
Numerical computations were carried out at the Supercomputer Center, 
the Institute for Solid State Physics, University of Tokyo. 
This work was supported by a Grant-in-Aid for Scientific Research from 
the Ministry of Education, Culture, Sports, Science, and Technology, Japan.

%

\end{document}